%% file: main.tex
\documentclass[conference]{IEEEtran}
\usepackage{graphicx}
\usepackage{epstopdf}
\usepackage{pgfplots}						
\pgfplotsset{compat=newest}
\usepackage{epstopdf}
\usepackage{tikz}
\pgfplotsset{compat=newest}
\pgfplotsset{plot coordinates/math parser=false}
\usepackage{standalone}
\usepackage{setspace}
\usepackage{float}
\usepackage{algorithm}
\usepackage{amsfonts}
\usepackage{ifthen}
\usepackage{booktabs}
\usepackage{amssymb} 
\usepackage{supertabular}
\usepackage{array}
\usepackage{multirow}
\usepackage{fancyhdr}
\usepackage{psfrag}
\usepackage{mathtools}
\usepackage{amsmath}
\usepackage{hhline}
\usepackage{type1cm}
\usepackage{lettrine}
\usepackage{cite}
\usepackage{fancybox}
\usepackage[hidelinks]{hyperref}
\usepackage{threeparttable}
\usepackage[english]{babel}
\usepackage{resizegather} 
\usepackage{epstopdf} 
\usepackage{array}
\usepackage{multirow}
\usepackage{booktabs}
\usepackage{url}
\usepackage{indentfirst}
\usepackage{comment}
\usepackage{placeins}
\usepackage{emptypage}
\usepackage{bm}
\usepackage{relsize,amsmath}
\usepackage{mathrsfs}  
\usepackage{enumitem}
\usepackage{lipsum}
\usepackage{blindtext}
\usepackage{cuted}
\usepackage{amsthm}
\usepackage[caption=false]{subfig}
\usepackage{siunitx}
\usepackage[normalem]{ulem}
\usepackage[short, nocomma]{optidef}
\makeatletter
\let\MYcaption\@makecaption
\def\BState{\State\hskip-\ALG@thistlm}
\makeatother

\makeatletter
\let\@makecaption\MYcaption
\makeatother

\theoremstyle{remark}

\theoremstyle{definition}

\usepackage{soul} 

\newcommand{\R}{\mathbb{R}}

\graphicspath{{Figures/}}

\usepackage{mfirstuc}

\addto\captionsenglish{}

\DeclareMathOperator{\diag}{diag}

\newcommand*{\C}{\mathbb{C}}

\newcommand*{\nPQ}{\textrm{n}}
\newcommand*{\idL}{\textrm{L}}
\newcommand*{\nall}{\textrm{N}}
\newcommand*{\nlines}{\textrm{m}}

\newcommand*{\cl}{\textrm{cl}}
\newcommand*{\fl}{\textrm{fl}}

\newcommand*{\rg}{\textrm{rg}}
\newcommand*{\cb}{\textrm{cb}}

\newcommand*{\conj}[1]{\overline{#1}} 

\usepackage{cleveref}
\crefname{equation}{}{}
\Crefname{equation}{Equation}{Equations} 
\crefname{table}{Table}{Tables}
\crefname{figure}{Fig.}{Fig.}
\Crefname{figure}{Figure}{Figures}
\crefname{section}{Section}{Sections}

\usepackage[nopostdot,nomain,numberedsection,nogroupskip]{glossaries}

\newglossary[alg]{acronym}{acr}{acn}{\acronymname}

\DeclareMathOperator*{\argmin}{arg\,min}

\makeatletter
\g@addto@macro\normalsize{%
	\setlength\belowdisplayskip{2pt}
	\setlength\belowdisplayshortskip{2pt}
}
\makeatother

\newglossarystyle{ieeenom}{%
\renewcommand{\glsgroupheading}[1]{}%
}
\newcommand{\mv}[1]{\boldsymbol{\mathbf{#1}}}

\newcommand{\sh}[1]{{\color{purple}#1}}

\newcommand\blfootnote[1]{%
  \begingroup
  \renewcommand\thefootnote{}\footnote{#1}%
  \addtocounter{footnote}{-1}%
  \endgroup
}
\begin{document} 


\title{On Distribution Grid Optimal Power Flow Development and Integration}

\author{\IEEEauthorblockN{Sarmad Hanif\IEEEauthorrefmark{1}, Rabayet Sadnan, Tylor E. Slay, Nawaf Nazir, \\Shiva Poudel, Bilal Bhatti, Andy Reiman, Jim Follum}
\IEEEauthorblockA{\textit{Paciﬁc Northwest National Laboratory} \\
Richland, WA, USA\\
\IEEEauthorrefmark{1}sarmad.hanif@pnnl.gov}
\and
\IEEEauthorblockN{Joseph McKinsey, Tarek Elgindy, Rui Yang}
\IEEEauthorblockA{\textit{National Renewable Energy Laboratory} \\
Denver, CO, USA}
}
\maketitle
\begin{abstract}
Due to changes in electric distribution grid operation, new operation regimes have been recommended. Distribution grid optimal power flow (DOPF) has received tremendous attention in the research community, yet it has not been fully adopted across the utility industry. Our paper recognizes this problem and suggests a development and integration procedure for DOPF. We propose development of DOPF as a three step procedure of 1) processing the grid, 2) obtaining a tractable solution, and 3) implementing multiple solution algorithms and benchmarking them to improve application reliability. For the integration of DOPF, we demonstrate how a DOPF federate may be developed that can be integrated in a co-simulation environment to mimic the real-world conditions and hence improve its practicality to be deployed in the field. To demonstrate the efficacy of the proposed methods, tests on IEEE 123-bus system are performed where the usage of tractable formulation in DOPF algorithm development and its comparison to the benchmark solution are demonstrated.
\blfootnote{This material is based on work supported by the U.S. Department of Energy (DOE). Pacific Northwest National Laboratory (PNNL) is operated for DOE by the Battelle Memorial Institute under Contract DE-AC05-76RL01830. Rabayet Sadnan was the intern at PNNL from Washington State University during the completion of this work.}
\end{abstract}

%
%
\section{Introduction}
\label{Sec:Introduction}

Electric distribution grids are experiencing significant changes in their infrastructure, such as installation of distributed energy resources (DERs). This infrastructure update must also accompany a renewed operation strategy for the distribution grids. Distribution grid optimal power flow (DOPF) could provide a pathway to implement such innovation in the distribution grids. In bulk power systems, optimal power flow (OPF) has received tremendous attention, as it has proved to dispatch flexible resources in order to balance supply and demand, while at the same time maintaining grid and device security constraints and limits~\cite{carpentier1962contribution}. Indeed, OPF is included in the most bulk power system Energy Management System (EMS) tools. However, DOPF is inherently different than OPF, as distribution grids have higher R/X ratio, have different phases in different portions of the grid and have a different topology. Hence, the same bulk system implementations are not simply transferable to DOPF and consequently to Distribution Management Systems (DMSs). 

The theoretical research on the DOPF formulation and solution algorithms is an active research area. Traditionally \textit{DistFlow} based algorithms that are derived from the branch power flow models are used to represent the DOPF problem~\cite{Baran.1989,nazir2020optimal}. To deal with the non-convexity of the \textit{DistFlow} equations, several methods have been proposed in the literature, including using linear approximations~\cite{Baran.1989}, convex relaxation methods~\cite{gan2014convex} and convex restriction methods~\cite{nazir2021grid}. In addition, in distribution systems, the scheduling of discrete mechanical devices such as transformers, switches and capacitor banks needs to be optimized in conjunction with continuous flexible loads~\cite{nazir2018receding}. 
With the emergence of these theoretical advancements, open-source example implementations of DOPF have been provided in the~\textit{PowerModels} package~\cite{coffrin2018powermodels} and the {\it{PowerModelsDistribution}}, specific for distribution systems~\cite{FOBES2020106664}. An OpenDSS driven OPF example is given in~\cite{rigoni2020open}.  

Even though DOPF algorithms and their examples are receiving theoretical attention, their inclusion in Advanced Distribution Management System (ADMS) platforms \cite{agalgaonkar2016adms} has still not taken place. Some examples of the state-of-the-art ADMS applications are Volt/Var control set-point generation and Conservative Voltage Reduction (CVR) mechanisms, which can be considered the derivatives of the possible DOPF problem formulations. Implementing a generalized DOPF solution to eradicate all possible issues in a distribution grid is a challenging task. This is because distribution grids vary greatly in terms of size, locations and their unique operational challenges (e.g., remote versus island grids). Hence, the one-size-fits all solution of bulk systems, which usually aim to minimize the total cost of energy, may not be easily transferable to distribution grids. As a result, the distribution grids realm requires additional research on efficient methods to design, develop, integrate and test DOPF algorithms and solutions. The present paper aims to address this critical gap in research and development.

This paper presents a step-by-step procedure for developing the DOPF solution and testing it in an integrated simulation environment. The paper demonstrates the core feature development procedure for the DOPF. This procedure is important because it helps the developer to implement a reliable DOPF solution. The paper proposes implementation is achieved by processing the grid models in 
 a form suitable for adoption by computationally efficient computer libraries/packages, developing approximation to reduce complexity and comparing alternate solutions. Co-simulation platforms are proposed to provide realistic testing conditions, since they allow for detailed physical simulations that mimic real-world communication infrastructure \cite{huang2018simulation}. Using the proposed core features, the paper demonstrates the development of a DOPF federate that can be integrated in a co-simulation environment. In the end, the paper presents an example of a fully integrated DOPF application in a co-simulation environment.

\input{DOPF_Development.tex}
\input{DOPF_Integration_Example.tex}
\input{case_studies}
\input{conclusion_outlook}
\input{appendix.tex}
\tiny{
\bibliographystyle{myIEEEtran}
\bibliography{abreviations_IEEE,library}
}

\end{document}

%% file: DOPF_Development.tex
\begin{figure}[h]
\centering
\includegraphics[width=0.45\textwidth]{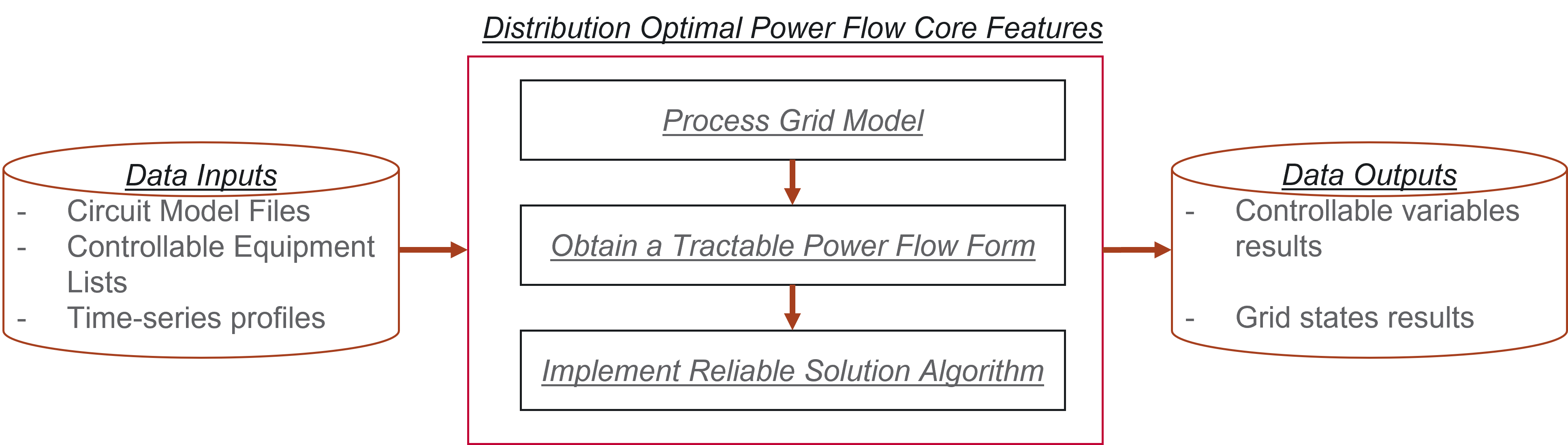}
\caption{Development of core DOPF Features.}
 \label{fig-flow-overview}
\end{figure}

\section{DOPF Development \& Integration}\label{overview}
In \cref{fig-flow-overview}, core features of DOPF development procedure are shown, along with the inputs required for and outputs generated from the successful demonstration of DOPF features. \\
\uline{\textit{Data Inputs and Outputs:}} For the distribution feeder of interest, its circuit model is the most important input for the DOPF development procedure, which allow to run power flows to validate that a viable solution of the DOPF exists. GridLAB-D and OpenDSS are common open source power flow simulators that can be utilized for this purpose. Other inputs include information on the equipment required to be controlled through DOPF. As penetration of DERs in distribution grids is increasing, multi-period optimization is of interest, and hence including time-series profiles can be considered essential for developing DOPF algorithms. From the core features of DOPF, expected outputs to be observed are the proposed changes in the controllable variables (e.g., taps positions, capacitor banks, controllable loads commands) and its impact on the grid states (e.g., node voltages and line currents).\\
\uline{\textit{DOPF Core Features Overview:}}
The core features of the DOPF development consist of 1) processing a grid model, 2) converting a highly non-convex distribution power flow problem into a tractable form, and 3) implementing an algorithm that can produce a reliable solution in terms of solution times and solution repeatability. The explanation of core features development and the rationale of the adopted steps in the procedure are presented next.
\subsection{DOPF Core Features Development}
\uline{\textit{Process Grid Model:}} The grid to be optimized by models needs to be converted to an appropriate vector and matrices format. This allows the DOPF developer to utilize packages/libraries which can improve computational performances (e.g., see Python's SciPy library implementation \cite{bressert2012scipy}). Based on the input circuit model format and the information available in the files, an appropriate topology reading algorithm may be written to develop such functionality as follows. First, an injection vector determining how much load/generation is injected/withdrawn at an appropriate node is developed. Second, incidence matrices mapping controllable equipment to the appropriate node in the injection vector is determined. Third, a verification model to map the injections to the grid states is developed to verify that the model obtained matches the real-world grid states. This step is basically translating the grid information into a form that is similar to power flow models, where admittance translates nodal power injections to voltage calculations. Consider a DOPF algorithm aiming to control load shedding $\mv{p}^{\fl}$, capacitor banks $\mv{q}_{\cb}$, tap-settings $ \mv{t}_{\rg}$, PV active power shed $\mv{p}_{\text{pv}}$ and PV reactive power injections $\mv{q}_{\text{pv}}$, and power feed in from the bulk power system $p_0/q_0$. The injection vector to be controlled is then:
\begin{align}
 \mv{u}=\{\mv{p}^{\fl}, \mv{q}_{\cb}, \mv{t}_{\rg}, \mv{p}_{\text{pv}}, \mv{q}_{\text{pv}}, p_0, q_0\}.
\end{align}
In this case, $\mv{x}$ is the set of state variables (like voltages and resultant current magnitude flows), $\mv{C_x}$ includes the impact of the controllable vector on state variables and $\mv{C_u}$ is the set of incidence matrices that maps the controllable equipment list onto the appropriate location in the network node list. An example of the incidence matrix for the regulator follows. An incidence matrix $\mv{C}_{\mv{r}}$ links regulator location to voltage changes such that $\mv{C}_{\mv{r}}(i,j)=1$ if tap regulator $j$ is upstream to node $i$, otherwise $\mv{C}_{\mv{r}}(i,j)=0$. The grid processing steps attempt to formulate the following equation to verify that the grid model is appropriate:
\begin{align}\label{eq:f}
 f(\mv{u}, \mv{x}, \mv{C_x}, \mv{C_u}) = \mv{0}
\end{align}
Different than the power flow problems, the DOPF requires non-equality constraints to be formulated too, such as voltage limitations, current flow limitations and controllable variables movement limitations. These constraints can be compacted in a similar manner to power flow model\footnote{where $\mv{D_x}$ is the inequality counterpart of $\mv{C_x}$, just as line flow utilizes different admittance quantities than nodal admittance.} as:
\begin{align}\label{eq:g}
 g(\mv{u}, \mv{x}, \mv{D_x}, \mv{C_u}) \leq \mv{0}
\end{align}

\uline{\textit{Obtain Tractable Power Flow Form:}} With the grid model processed and verified, the next step of the DOPF core feature development is to convert the non-convex distribution power flow model, along with the discrete controllable equipment functional space, into a tractable model form. The reason for performing this step is that distribution grids differ greatly in design and in the spirit of standardizing DOPF model development; a tractable power flow form must be made that can achieve a balance between the solution space and computational efforts. As a case of the simplest tractable form, a linear power flow model with continuous decision variables mapping can be picked. To ensure the tractable model form is able to be used in developing the DOPF algorithm, the model should be parameterized in the controllable equipment variables. This allows for validation of the errors on the approximation of the power flow as the change in the controllable equipment is proposed by the DOPF solution. From the most general grid model described above, a tractable approximation (e.g., linear model) can be expressed as:
\begin{align}\label{eq:lin_model}
    \tilde{\mv{x}} &= \mv{A}\mv{u} + \mv{C_u}\mv{u}.
\end{align}
where the above model is the linear mapping of control variables $\mv{u}$ to approximated grid states $\tilde{\mv{x}}$ using a linear sensitivity matrix $\mv{A}$ and the incidence matrix $\mv{C_u}$.
\vspace{-0.1cm}
\uline{\textit{Implement Reliable Solution Algorithm:}} This step utilizes the processed grid model and its approximation to develop a reliable solution format. By ``reliable,'' it is meant that the solution algorithm should provide acceptable results in a timely fashion. This can be achieved by implementing a solution algorithm using a standard off-the-shelf solver, which can then be benchmarked against the individual algorithms developed by the user. This step is more than just converting power flow formulation to an optimization problem. This step is introduced in the DOPF core features to allow for testing various different algorithms and techniques. These are necessary as a wide number of distribution grid operation techniques can be developed depending on the application and necessity of the operator, and such a solution algorithm development process can be used to obtain a reliable solution algorithm. Next, we present two examples where an approximate model \cref{eq:lin_model} is shown to be utilized for setting up 1) a benchmark method that utilizes a standard DOPF formulation to be solved by an off-the-shelf solver and 2) an alternate method that solves the standard DOPF formulation using a different algorithm. 

\begin{mini!}|s|[2]
    {}{\mv{c}\big(\mv{u},\tilde{\mv{x}} \big) \label{eq:apprx_DSOprob_OF_1}}
    {\label{eqs:DSOprob_1}}
    {}\addConstraint{\mv{f} (\mv{u},\tilde{\mv{x}})  }{= 0   \label{eq:DSOprob_b_1}}{:\mv{\lambda}}
    \addConstraint{\mv{g} (\mv{u},\tilde{\mv{x}})  }{\leq 0   \label{eq:DSOprob_b_2}}{:\mv{\mu}}
\end{mini!}
where $\mv{c}(\cdot)$ is the objective function term to be minimized and \cref{eq:DSOprob_b_1} and \cref{eq:DSOprob_b_2} are the approximated counterparts of \cref{eq:f} and \cref{eq:g}, formulated in a similar manner as \cref{eq:lin_model}. Variables $\mv{\lambda}$ and $\mv{\mu}$ are the Lagrange multiplier of the respective constraints. To compliment the tractability of the optimization problem, similar to approximation of constraints, the objective function can also be made linear or quadratic. 
The standard  DOPF formulation of \cref{eqs:DSOprob_1} could be solved using an off-the-shelf solver \cite{GurobiOptimization.2016}. However, below is an example of an alternate solution algorithm, which can be developed and benchmarked against \cref{eqs:DSOprob_1}. Consider the Lagrangian of \cref{eqs:DSOprob_1}:
\begin{align}
    &\mathcal{L}(\mv{u},\tilde{\mv{x}}, \mv{\lambda}, \mv{\mu})
    = \mv{c}\big(\mv{u},\tilde{\mv{x}}\big) - \mv{{\lambda}}(\mv{f} (\mv{u},\tilde{\mv{x}})) 
     -{\mv{\mu}}(\mv{g} (\mv{u},\tilde{\mv{x}})),
\end{align}
which can be deployed to update primal and dual variables at each iteration $k$ using the following rules:
\begin{subequations}
\begin{align}\label{eq:algorithmupdate_1}
     \text{Primal Updates: \ \ } \mv{u}^{k+1} &= \argmin \mathcal{L}_{\mv{u}}(\cdot)  \\
     \mv{x}^{k+1} & = \argmin \mathcal{L}_{\mv{x}}(\cdot) \\  
     \text{Dual Updates: \ \ } \mv{\lambda}^{k+1} &= \mv{\lambda}^{k} - \alpha\Delta \mathcal{L}_{\mv{\lambda}}(\cdot) \\ 
    \mv{\mu}^{k+1} &= \mv{\mu}^{k} - \beta \Delta \mathcal{L}_{\mv{\mu}}(\cdot),
\end{align}
\end{subequations}
until there are no variables updates. In \cref{eq:algorithmupdate_1}, $\Delta \mathcal{L}_{\cdot}(\cdot)$ is the first order derivative of $\mathcal{L}$ with respect to $(\cdot)$ and $(\alpha, \beta)$ are small positive scalars. The algorithm \cref{eq:algorithmupdate_1} is formally known as the gradient descent \cite{Boyd.2004}.

\vspace{-0.5cm}
\begin{figure}[h]
\centering
\includegraphics[width=0.4\textwidth]{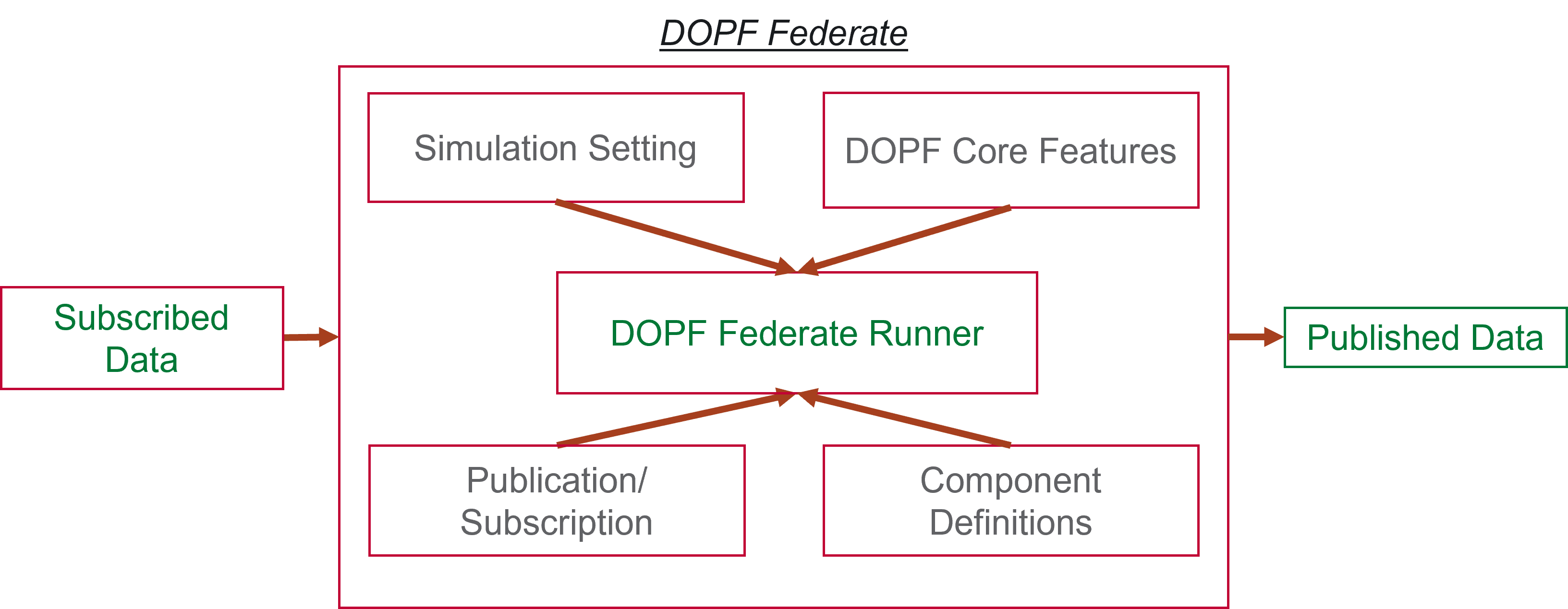}
\caption{DOPF federate components}
 \label{fig-dopffedrun}
\end{figure}

\subsection{DOPF Federate Development}\label{sec-dopf-fed}

Co-simulation environments have been proposed for realistic assessment of the developed applications and services for the grid \cite{palmintier2017design}. In these co-simulation environments, each individual simulation environment is called a federate, which has a pre-defined set of inputs, outputs and simulation settings. This federate architecture is particularly inspired by the co-simulation framework called Hierarchical Engine for Large-Scale
Infrastructure Co-Simulation (HELICS) \cite{palmintier2017design}. 
\Cref{fig-dopffedrun} shows components of the DOPF federate runner. In \cref{fig-dopffedrun}, basic structure of the DOPF federate is shown where it is set up to subscribe data from and publishes data to other federates.\\
 \uline{\textit{Simulation Setting:}} These settings include simulation duration, update intervals and specification on the types of core features to be selected for the federate.\\
 \uline{\textit{Publication/Subscription:}} These contain information on the data to be inputted (subscribed) from and outputted (published) to relevant feeders to achieve the desired DOPF federate functionality. For example, current tap positions of the installed regulators would be required to calculate an optimal setpoint from the DOPF.\\
 \uline{\textit{Component Definitions:}} These definitions decompose the publication and subscription requirements into frequency (e.g., static versus dynamic), data types (e.g., normal versus sparse matrices) and appropriate federate mappings (e.g., control variables to be sent to the feeder and current state to be obtained from the appropriate sensors). These definitions also allow to switch out the relevant federates so that different testing environments may be created (e.g., removing state estimator and replacing with actual quantities from feeder federate etc.) for the DOPF federate. \\
\uline{\textit{DOPF Federate Runner:}} The runner script implements the DOPF components such that it is able to 1) demonstrate the required DOPF functionality (e.g., ensuring calling the appropriate DOPF feature) and 2) is coordinated with other federates, i.e., stopping the DOPF federate at the appropriate times to input and output the data (e.g., generating control setpoints every 30 minutes and inputting voltage estimates every 25 minutes so that the DOPF functions are executed within the required time.)

%% file: DOPF_Integration_Example.tex
\vspace{-0.5cm}
\begin{figure}[h]
\centering
\includegraphics[width=0.35\textwidth]{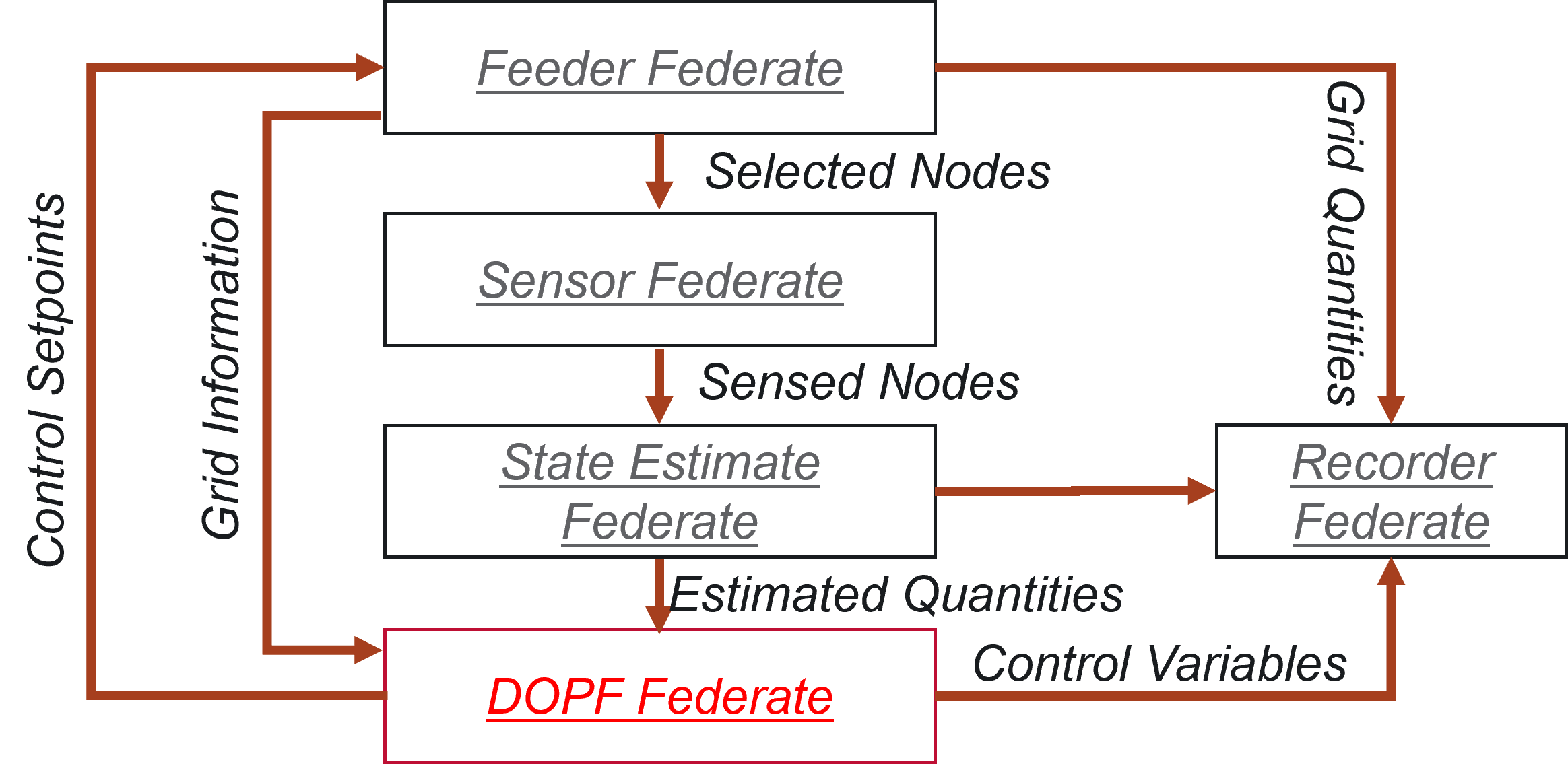}
\caption{Example DOPF Integration in a Co-Simulation Environment}
 \label{fig-flow-overview2}
\end{figure}
\vspace{-0.5cm}
\subsection{DOPF Federate Integration}\label{sec-dopf-integration}
This section presents an example integration of the DOPF federate in a co-simulation environment. In \cref{fig-flow-overview2}, ``quantities'' are referred to complex nodal voltages and powers.\\
\uline{\textit{Feeder Federate:}} This represents an abstraction of a real-world grid in a co-simulation environment, which simulates the distribution grid power flow and generates appropriate quantities to be treated as the ``ground-truth.'' \\
\uline{\textit{Sensor Federate:}} For practicality purposes, information on the limited and practical sensor equipment of the distribution grid is simulated using a sensor federate. Inputs from the feeder federate on the nodes to be sensed are provided to this federate. Either less than total nodes, or noise on the ``ground truth'' quantities are assumed by the sensor federate to demonstrate the practical constraints of the grid.\\
\uline{\textit{State Estimate Federate:}} From the sensed quantities, the state estimator federate estimates the grid quantities.\\
\uline{\textit{Recorder Federate:}} To analyze performances of the federates, recorder federate is used to record the relevant quantities.\\
\uline{\textit{DOPF Federate:}} From the estimated grid quantities and grid topology information, the DOPF calculates control setpoints to be passed to the feeder federate. Note that as the model based DOPF core features are in discussion in this paper, grid information (e.g., topology and location of controllable equipment) is required. However, as data-driven (model-free) approaches are proposed in the literature, such requirements may be relaxed and hence may improve the overall data requirements for testing and developing DOPF federate. The DOPF federate runner described in \cref{sec-dopf-fed} coordinates the timing between different federates, i.e., obtaining relevant information (grid information and estimated quantities) and generating control setpoints in a timely fashion. Similar to the state estimate federate and feeder federate, variables determined by the DOPF federate are saved by the recorder federate. \Cref{sec:gadal-integration} gives an overview of the above-mentioned DOPF development and integration implementation \cite{Hanif_Distributed_Optimal_Power_2022}.

%% file: case_studies.tex
\begin{figure}[h]
\centering
\includegraphics[width=0.40\textwidth]{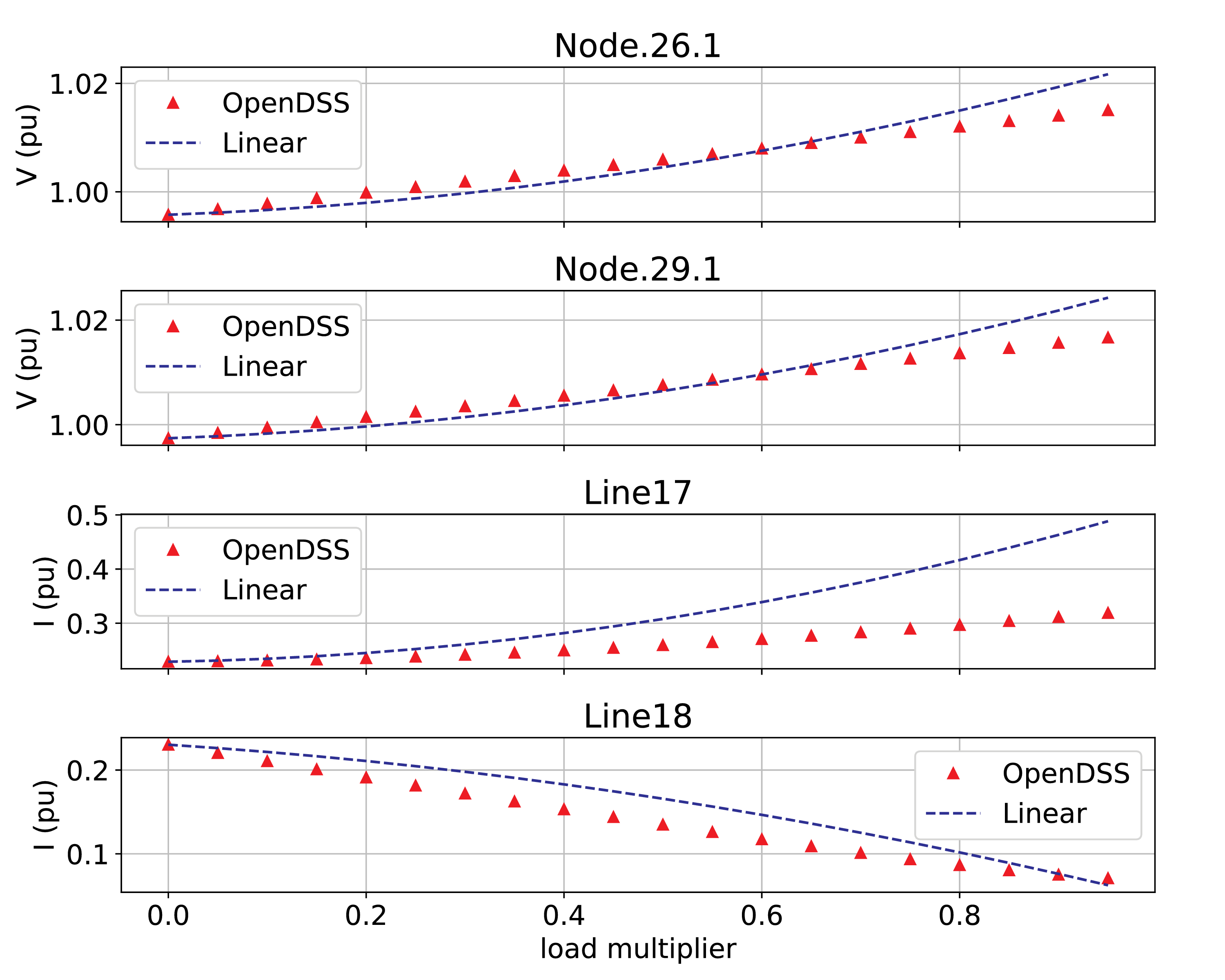}
\caption{Linearization Results of 123-Bus System.}
 \label{fig:Volt_flow_linearizations_123}
\end{figure}
\begin{figure}[h]
\centering
\includegraphics[width=0.40\textwidth]{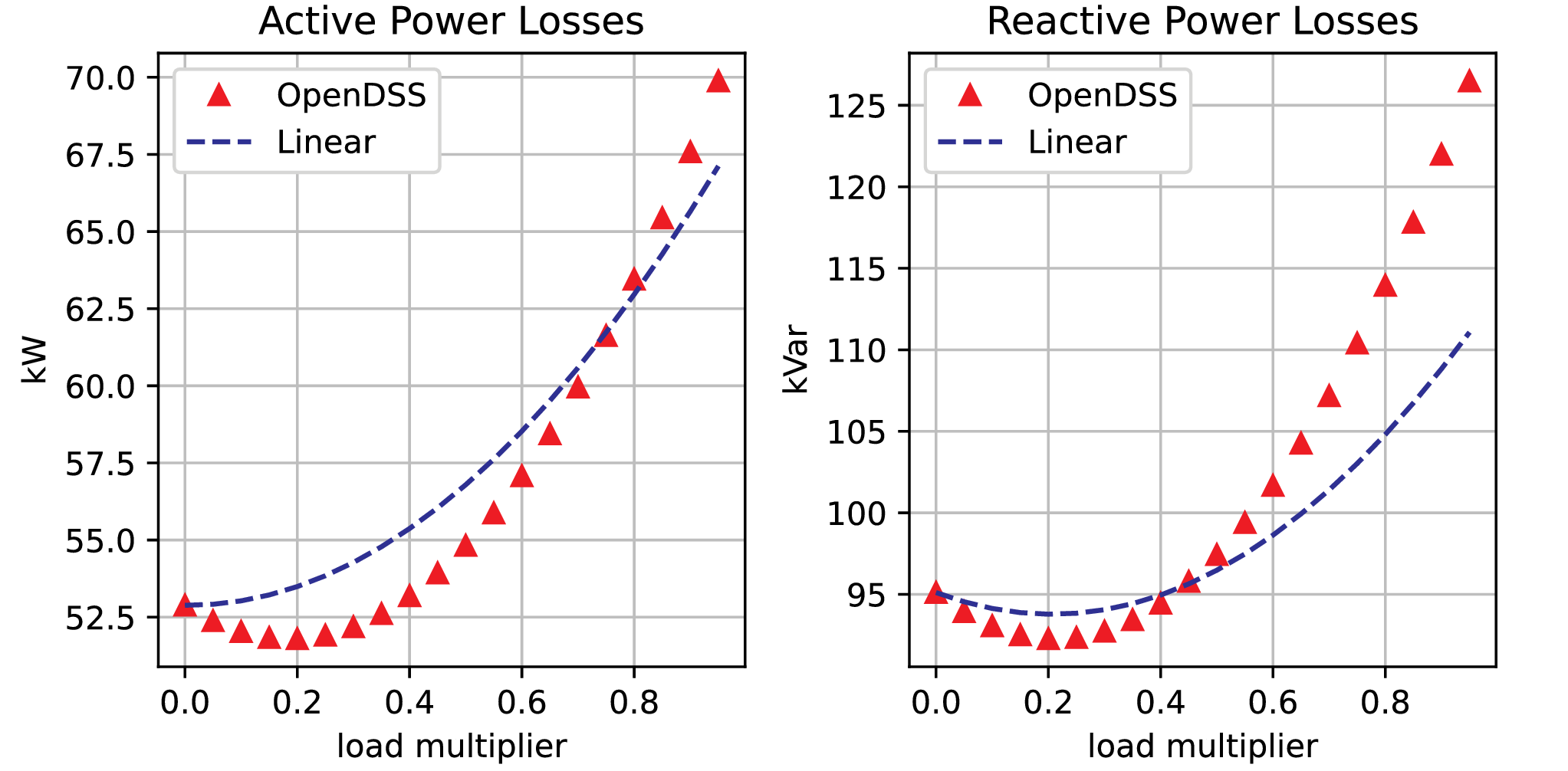}
\caption{Power Loss Linearization Results for 123-Bus.}
 \label{fig:Loss_linearization_123}
\end{figure}
\section{Case Studies and Analysis}\label{Sec:Results}
\begin{figure}[ht]
\centering
\includegraphics[width=0.5\textwidth]{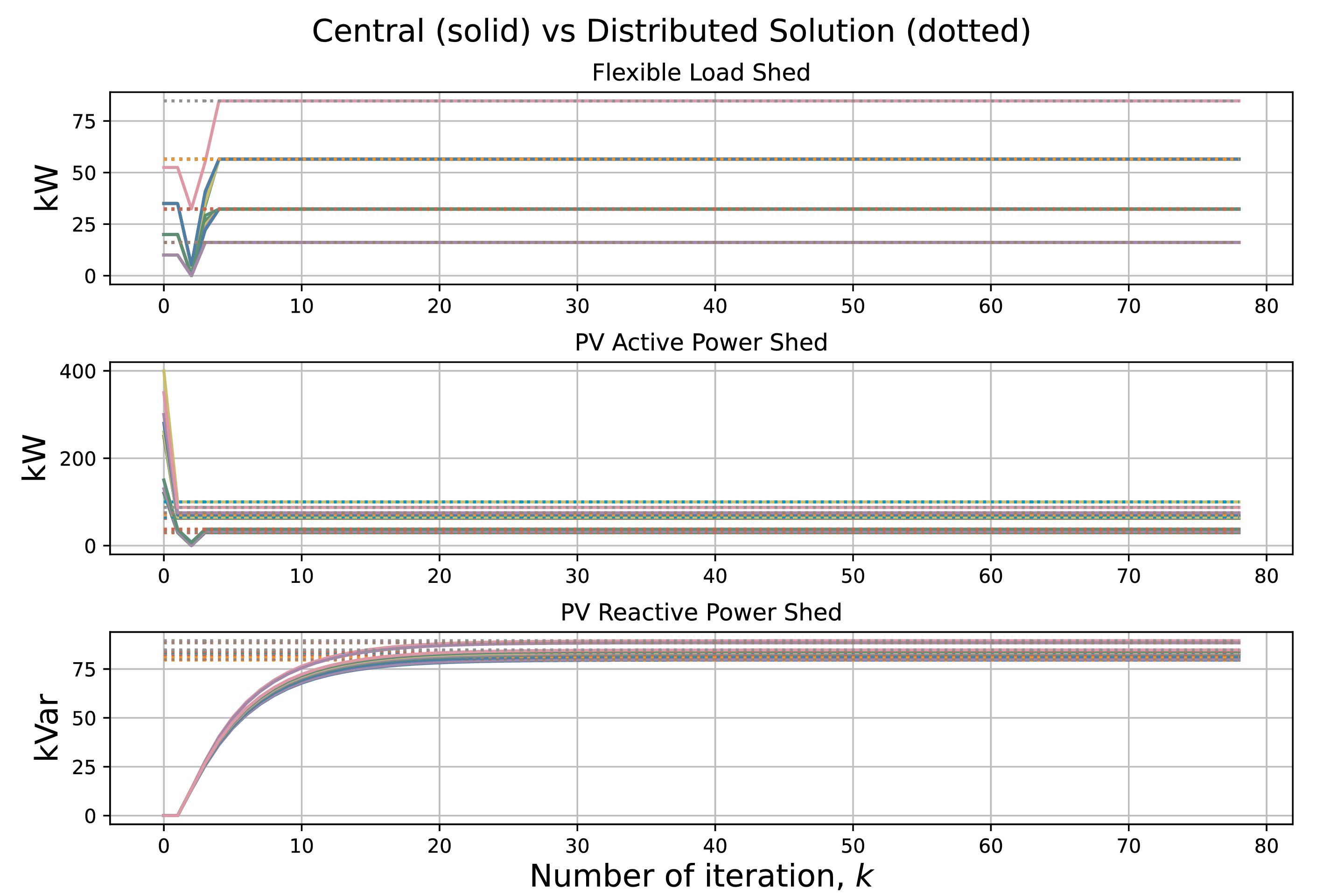}
\caption{Benchmarking Distributed Solution across Iterations with the Central Solution Obtained from an Off-the-Shelf Solver for 123-Bus.}
 \label{fig:Benchmark_opf_123}
\end{figure}

\begin{figure}[h]
\centering
\includegraphics[width=0.5\textwidth]{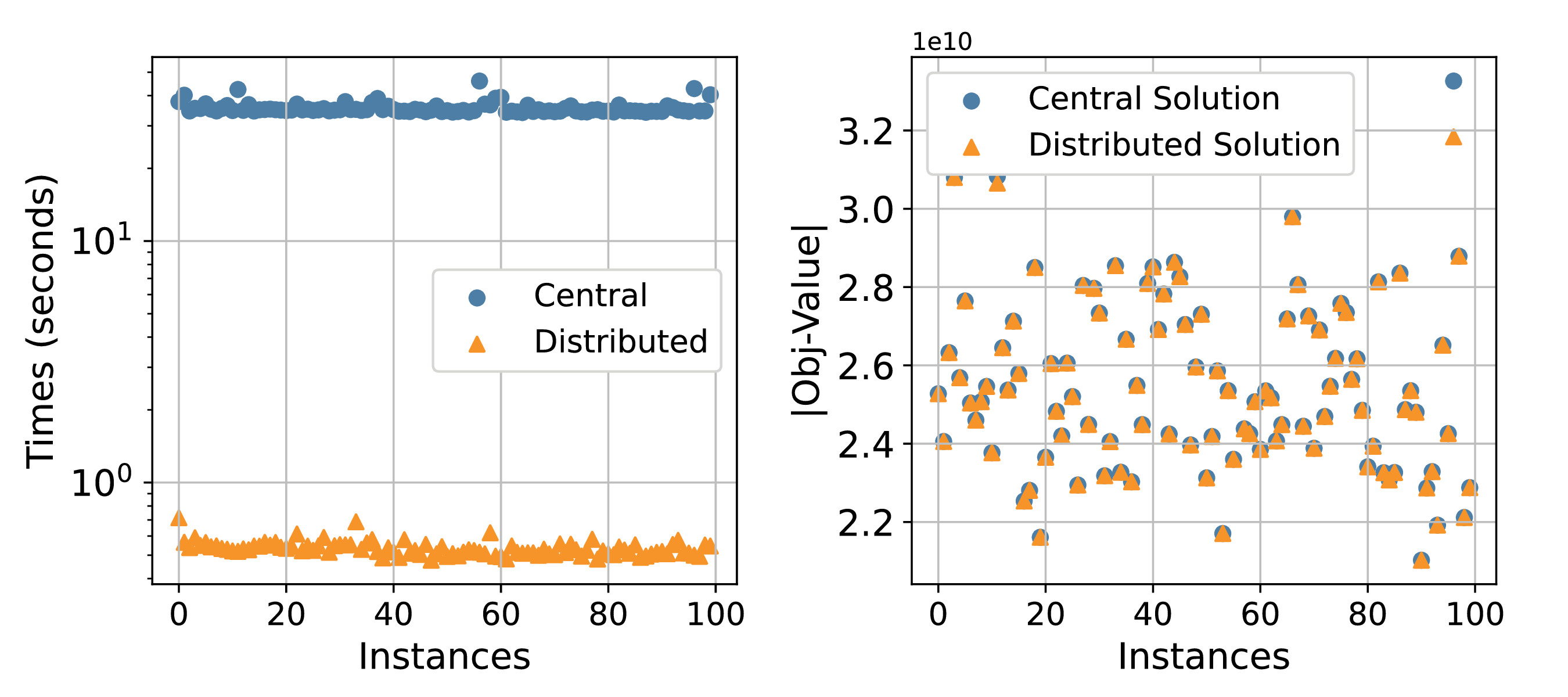}
\caption{Statistics of 100 Randomly Initialized DOPF problem.}
 \label{fig:CvD_Stat}
\end{figure}

This section presents results from the implemented DOPF core features. For the simulation, we have used the modified IEEE 123-bus test system. First we validate the linear power model, and then we compare the own implemented gradient-descent based distributed DOPF solution with its central variant which uses the off-the-shelf solver \cite{GurobiOptimization.2016} to showcase the scalability and computational efficiency.\\
\uline{\textit{Linear Model Analysis:}}
First, we showcase the validation of the linearized power flow model by changing the (i) flexible load, (ii) PV active power generations, (iii) regulator taps and (iv) kVars of the capacitor banks by introducing a multiplier -- denoted as \textit{``Load Multiplier''} -- and changing the value of the variable \textit{``Load Multiplier''} $\in [0,~1]$. Then we compare the nonlinear nodal voltage, current in the lines and line losses with approximated linear solutions, where OpenDSS is used to obtain the nonlinear solutions \cite{openDSS.2018}. In Fig. \ref{fig:Volt_flow_linearizations_123}, the linear approximated voltage and current for different nodes and lines are shown, where it can be seen relatively accurate linearized model is obtained around which can be used to generate DOPF solutions.
In addition to that, the approximated line losses have been compared in Fig. \ref{fig:Loss_linearization_123}, a common feature in DOPF minimization objectives. Similar to the node voltage and line current approximations, both the total kW and total kVar line losses can be approximated with the linearized model.\\
\uline{\textit{Solution Benchmark:}}
We benchmark the distributed DOPF solution (implemented using gradient descent algorithm) with its centralized variants implemented using off-the-shelf solver. We utilize (i) flexible loads, (ii) PV active power sheds, (iii) PV reactive power sheds, (iv) regulator tap settings and (v) capacitor banks switches for the optimal operation of the system. The problem objective is to reduce the cost of operation by minimally changing the current setpoints when subjected to a sudden change in the system and bring the system to a state in which the pre-specified operational limit has been maintained. The benchmark is shown in Fig. \ref{fig:Benchmark_opf_123} and Fig. \ref{fig:CvD_Stat}. We compare the major optimal control variables -- the shed of flexible loads and PV power outputs in Fig. \ref{fig:Benchmark_opf_123}; distributed control variables reaches the central solution in 79 iterations. Other control variables, such as cap bank switches and tap positions, remain the same for the system changes and thus produce trivial solutions, which are therefore not presented in this paper.
To demonstrate that the DOPF solution is reliable and scalable, we create 100 different instances of control variable changes and compare both the distributed with central solutions. In the left figure of Fig. \ref{fig:CvD_Stat}, we see that the distributed solution is almost 100 times faster than the central solution -- $\sim$ 0.4 seconds  compared to $\sim$40 seconds -- and thus scalable for larger power distribution systems. However, the solution quality is not compromised as the objective value matches the central solutions (see the right figure of Fig. \ref{fig:CvD_Stat}). 
 

%% file: conclusion_outlook.tex
\vspace{-0.5cm}
\section{Conclusion}\label{Sec:Conclusions}
This paper presented a DOPF development and integration method. The intention of the paper was to help bring DOPF from a research tool to a more widely accepted solution in the industry. To that end, the paper presented a systematic way of developing and integrating a DOPF solution. For the development of DOPF, the three steps of a reliable solution development were: 1) processing the grid, 2) obtaining a tractable solution, and 3) implementing multiple solution methodologies for benchmarking purposes. Tests were conducted on IEEE 123-bus system where multiple solution algorithms implementation allowed reliably testing the developed DOPF scheme.

%% file: appendix.tex
\vspace{-0.3cm}
\section{Appendix}\label{sec:gadal-integration}
\begin{figure}
    \centering
\includegraphics[width=0.5\textwidth]{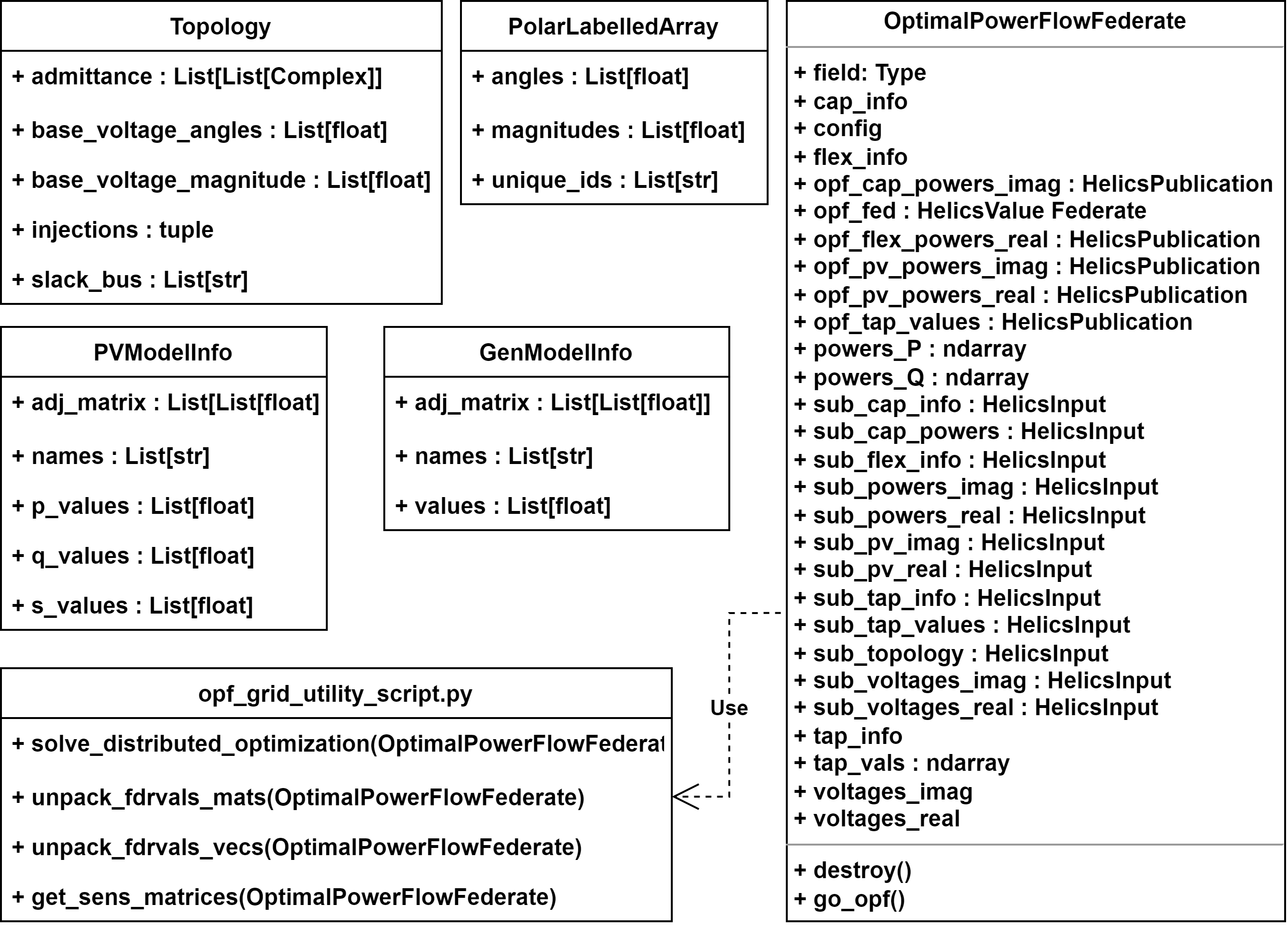}
    \caption{UML Diagram for the DOPF Federate}
    \label{fig:dopf-uml}
\end{figure}

\begin{figure}[h]
    \centering
    \includegraphics[width=0.25\textwidth]{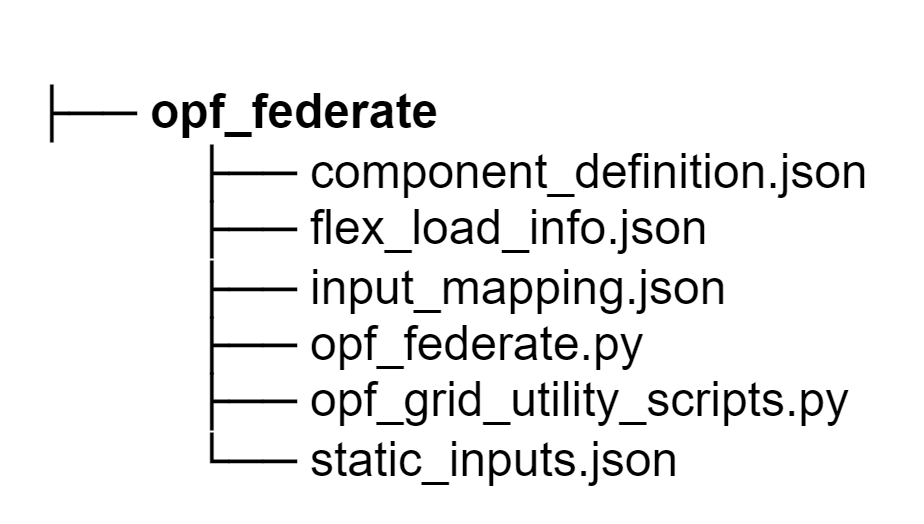}
    \caption{DOPF Implementation File Structure}
    \label{fig:dopf-tree}
\end{figure}

The DOPF integration example implementation is shown in Fig. \ref{fig:dopf-tree}. 
Each federated component must have its own directory with a minimum of the \texttt{component\_definitions.json} and the Python script to execute. The component definition file points the federate to the file to execute, maps the static inputs, and dynamic inputs/outputs. The DOPF implementation requires a \textit{scenarios} directory to initiate setup for the \texttt{LocalFeeder} models. The \texttt{LocalFeeder} directory generates the required DOPF information from the specified model and runs the co-simulation used by the DOPF federate. The \texttt{measuring\_federate} instantiates the federate variables to be published and \texttt{recorder} instantiates the federate variables to be subscribed to. The \texttt{wls\_federate} subscribes to the model voltages and powerflow for the given topology and publishes the estimated voltages. Figure \ref{fig:dopf-uml} outlines the DOPF federate and the primary utility functions used to solve. The \textit{OptimalPowerFlowFederate} class is primarily used to store the model information and establish the designated variable subscriptions and publications. The \texttt{opf\_grid\_utility\_script.py} contains the methods for federate matrices and vector ingestion. 

